\documentclass[aps,prd,onecolumn,amssymb,amsmath,nofootinbib,floatfix,notitlepage]{revtex4-1}

\usepackage{graphicx}
\usepackage{caption}

\newcommand{\p}{\ensuremath{\partial}}
\newcommand{\del}{\ensuremath{\delta}}
\newcommand{\Del}{\ensuremath{\Delta}}
\newcommand{\eps}{\ensuremath{\varepsilon}}
\newcommand{\ep}{\ensuremath{\epsilon}}
\newcommand{\tht}{\ensuremath{\theta}}
\newcommand{\lam}{\ensuremath{\lambda}}
\newcommand{\Lam}{\ensuremath{\Lambda}}
\newcommand{\sig}{\ensuremath{\sigma}}

\newcommand{\Om}{\ensuremath{\Omega}}

\newcommand{\Msun}{\ensuremath{M_{\odot}}}
\newcommand{\etal}{\emph{et al}.}

\newcommand{\Cite}[1]{Ref. \cite{#1}}
\newcommand{\Cites}[1]{Refs. \cite{#1}}
\newcommand{\eqn}[1]{Eqn. \eqref{#1}}
\newcommand{\eqns}[1]{Eqns. \eqref{#1}}
\newcommand{\fig}[1]{Fig. \ref{#1}}

\newcommand{\ph}[1]{\phantom{#1}}
\newcommand{\tab}[1]{Table \ref{#1}}

\newcommand{\be}{\begin{equation}}
\newcommand{\ee}{\end{equation}}
\newcommand{\la}[1]{\label{#1}}

\newcommand{\Cal}[1]{\ensuremath{\mathcal{#1}}}

\newcommand{\avg}[1]{\ensuremath{\langle \,#1\, \rangle}}
\newcommand{\delc}{\ensuremath{\delta_c}}
\newcommand{\fnl}{\ensuremath{f_{\rm NL}}}

\newcommand{\lamint}[1]{\ensuremath{\int_{-\infty}^{\infty}{\frac{d\lam}{\sqrt{2\pi}} #1}}} 

\begin{document}
\title{The Extreme Tail of the Non-Gaussian Mass Function}
\author{Aseem Paranjape}
\email{aparanja@ictp.it}
\affiliation{Abdus Salam ICTP, Strada Costiera 11, 34151 Trieste, Italy\\}
\author{Christopher Gordon}
\email{cxg@astro.ox.ac.uk}
\affiliation{Oxford Astrophysics, Physics, DWB, Keble Road, Oxford,
  OX1 3RH, UK\\}
\author{Shaun Hotchkiss}
\email{shaun.hotchkiss@helsinki.fi}
\affiliation{Department of Physics, University of Helsinki and
  Helsinki Institute of Physics, P.O. Box 64, FIN-00014 University of
  Helsinki, Finland} 
\begin{abstract}
\noindent
Number counts of massive high-redshift clusters provide a window to
study primordial non-Gaussianity. The current quality of data,
however, forces the statistical analysis to probe a region of
parameter space -- the extreme tail of the mass function -- which is
neither accessible in any of the currently available theoretical
prescriptions for calculating the mass function, nor calibrated in
$N$-body simulations. In this work we present a new analytical
prescription for calculating a ``resummed'' non-Gaussian halo mass
function, which is constructed to remain stable in the extreme
tail. We show that the prescription works well in the parameter regime
that has been currently explored in simulations. We then use Fisher
matrix techniques to compare our prescription with an extrapolated fit
to $N$-body simulations, which has recently been used to obtain
constraints from data collected by the South Pole Telecope. We show
that for the current data, both prescriptions would lead to
statistically consistent constraints. As the data improve, however,
there is a possibility of introducing a statistically significant bias
in the constraints due to the choice of prescription, especially if
non-Gaussianity is scale dependent and becomes relatively large on
cluster scales. It would then be necessary to test the accuracy of the
prescriptions in $N$-body simulations that can probe clusters with
high masses and redshifts in the presence of large non-Gaussianity. 
\end{abstract}
\maketitle
\section{Introduction}
\noindent
One of the key aims in cosmology  is to determine
how much non-Gaussianity there was  in the
primordial curvature fluctuations. Canonical single
field inflation with a smooth potential predicts that primordial
non-Gaussianity (NG) should be extremely small
\cite{Maldacena:2002vr,Acquaviva:2002ud}. Detecting\footnote{See
  \Cite{Desjacques:2010jw} for a review.} an appreciable 
amount of primordial NG could therefore rule out these
models. Measurements of the cosmic microwave background (CMB) 
\cite{Komatsu:2010fb}  and the scale-dependent halo bias
\cite{Slosar:2008hx}  provide good constraints on primordial
NG on scales of about $k\sim 0.05h^{-1}$Mpc. Although the
\emph{Planck} experiment will improve the range of scales constrained 
\cite{LoVerde:2007ri,Sefusatti:2009xu}, the primary CMB will alway be limited  
by foregrounds to $k\lesssim 0.2h^{-1}$Mpc. Since it is possible for
primordial NG to be quite strongly scale dependent
\cite{Riotto:2010nh,Byrnes:2009pe}, it is important to constrain it on
as wide a range of scales as possible, with complementary probes when 
possible. Number counts of galaxy clusters provide such a probe on
smaller scales \cite{LoVerde:2007ri} (see also
\Cites{Cunha:2010zz,Sartoris:2010cr,Fedeli:2010ud}). There have been
several recent attempts at using massive high-redshift clusters to constrain
primordial NG \cite{Jimenez:2009us, Cayon:2010mq,
  Hoyle:2010ce,Enqvist:2010bg,Williamson:2011jz}. 
Williamson \etal\  (W11) \cite{Williamson:2011jz} analysed a subset of
the clusters detected by the South Pole Telescope (SPT), using a
likelihood analysis which carefully accounts for issues such as the
survey selection function. 
They find a  posterior probability distribution for the standard
non-Gaussianity parameter \fnl, and their quoted result is
$\fnl=20\pm450$ at $68\%$ confidence. 

A key ingredient in any such
analysis is the chosen prescription for the non-Gaussian halo mass
function. Given a mass function, one can construct a likelihood for
the data by computing the expected number of clusters in a given mass
and redshift range, as an integral of the mass function. The W11
analysis uses the prescription of Dalal \etal\ (D08)
\cite{Dalal:2007cu}, which is essentially a fit to $N$-body
simulations.  As we will see however, the W11 likelihood function
probes a region of parameter space (the extreme tail of the mass
function) which has not been calibrated by D08 (nor indeed, by anyone
else). Additionally, \emph{all} the other currently available
prescriptions \cite{Matarrese:2000iz, LoVerde:2007ri, Maggiore:2009rx,
  D'Amico:2010ta, LoVerde:2011iz} formally break down in this extreme
region. 

It is therefore interesting to ask whether one can obtain any
analytical understanding of the mass function in its extreme tail, and 
whether the D08 prescription might be introducing a bias in the
analysis. The main result of this work is a new prescription for the
non-Gaussian mass function, which involves a resummation of an
infinite perturbative series. The resulting expression compares well
with the results of $N$-body simulations. More importantly, it
does not formally break down in the region of parameter space that
needs to be probed by the W11 data analysis, and can therefore be used
to compare against the D08 prescription. Using a Fisher
analysis, we find that for the current quality of data the two
prescriptions give statistically identical results. With better quality
data however, we show that there can be a statistically
significant bias between the two methods, especially if \fnl\ on small
scales is large. It would then be necessary to accurately calibrate
the tail of the non-Gaussian mass function in simulations.

The paper is organised as follows : Section \ref{extail} motivates and
introduces our new ``resummed'' prescription for the mass function,
comparing it with other prescriptions, both in regimes which have been
tested by simulations and in the extreme tail which has not. In
section \ref{fisher} we use a Fisher analysis to compare the error bars on
\fnl\ from the D08 and resummed prescriptions, and analyse the level of
bias between the two methods. We end with a brief discussion in
section \ref{conclude}. Technical details of calculations have
been relegated to the Appendix. Unless otherwise specified, we assume
a flat \Lam CDM cosmology with parameter values compatible with WMAP7 
\cite{Komatsu:2010fb} :  $h=0.703$ with $H_0=100h{\rm kms}^{-1}{\rm 
  Mpc}^{-1}$ the Hubble constant, total matter density $\Om_m
h^2=0.134$, baryonic matter density $\Om_b h^2=0.0227$, scalar
spectral index $n_s=0.966$, and $\sigma_8=0.809$. Throughout we
use the transfer function of Bardeen \etal\ \cite{Bardeen:1985tr},
with a baryonic correction as prescribed in \Cite{Sugiyama:1994ed}. 

\section{Number Counts in the Extreme Tail}
\la{extail}
\noindent
Within the paradigm of hierarchical structure formation, any
physically acceptable mass function\footnote{By mass function we mean
  $f_{\rm sky}(dV/dz)(dn/dM)$; where $dn$ is the comoving number density
  of halos with masses in $(M,M+dM)$, $f_{\rm sky}$ is the fraction of
  sky observed and $dV$ is the volume element with $dV/dz = 4\pi
  H(z)^{-1} \left[\int_0^zdz'H(z')^{-1}\right]^2$.} must be compatible
with the fact that massive objects tend to form late in the cosmic
evolution. Mass functions derived assuming Gaussian  initial
conditions
\cite{Press:1973iz,Bond:1990iw,Sheth:1999mn}
 easily
fit the bill by virtue of being monotonically decreasing at large
masses and redshifts. The situation with  non-Gaussian initial
conditions is not as clean however. There are several prescriptions in
the literature for theoretically calculating the non-Gaussian mass
function \cite{Matarrese:2000iz, LoVerde:2007ri, Maggiore:2009rx, 
  D'Amico:2010ta, LoVerde:2011iz}.  All of these rely on a
perturbative treatment of non-Gaussianity and assume a condition which
is at least as strong as $|\ep\nu|<1$ (see D'Amico \etal\ (D11)
\cite{D'Amico:2010ta} for a detailed comparison), where
$\ep\sim\fnl\sqrt A$ is a small parameter controlling the
non-Gaussianity ($A\sim10^{-9}$ being the power spectrum
normalisation) and 
\be
\nu(M,z)\equiv \frac{\delc}{\sig_M}\frac{D(0)}{D(z)}\,,
\la{nu-def}
\ee
with the threshold $\delc=1.686$ (for spherical collapse, but see
below), the linear growth rate $D(z)$, and the variance
$\sig^2_M\equiv\avg{\hat\del_M^2}$ of the linearly extrapolated
density field smoothed on scale $R=(3M/4\pi\bar\rho)^{1/3}$.  While
several of these mass functions have been tested in $N$-body
simulations and do reasonably well in their regime of validity, when
extrapolated to $|\ep\nu|>1$ they lead to unphysical results (e.g. the
Matarrese \etal\ (MVJ) \cite{Matarrese:2000iz} result becomes
imaginary for $\fnl>0$ when $\ep\nu\sim3$). This would be an academic
issue, were it not for the fact that the current quality of data
requires predictions of number counts at combinations of $(\fnl,M,z)$
values which lie squarely in the $|\ep\nu|>1$ corner
\cite{Enqvist:2010bg}. To see this, note that e.g. the W11 analysis
proceeds by calculating the joint probability, for a given \fnl, of
observing the ensemble of clusters in their sample, \emph{and} of
observing nothing else in the parameter range they explore. The large
errors on \fnl\ ($\sim450$) indicate that their likelihood is
non-negligible at say $\fnl\lesssim750$.  \fig{fig-epnu} shows
$\eps_1\nu$ as a function of mass at redshift $z=1.5$ for some
representative \fnl\ values, where   
\be
\eps_1 \equiv \avg{\hat\del_M^3}/\sig_M^3\,,
\la{eps1-def}
\ee
\begin{figure}[t]
\centering
\includegraphics[height=0.3\textheight]{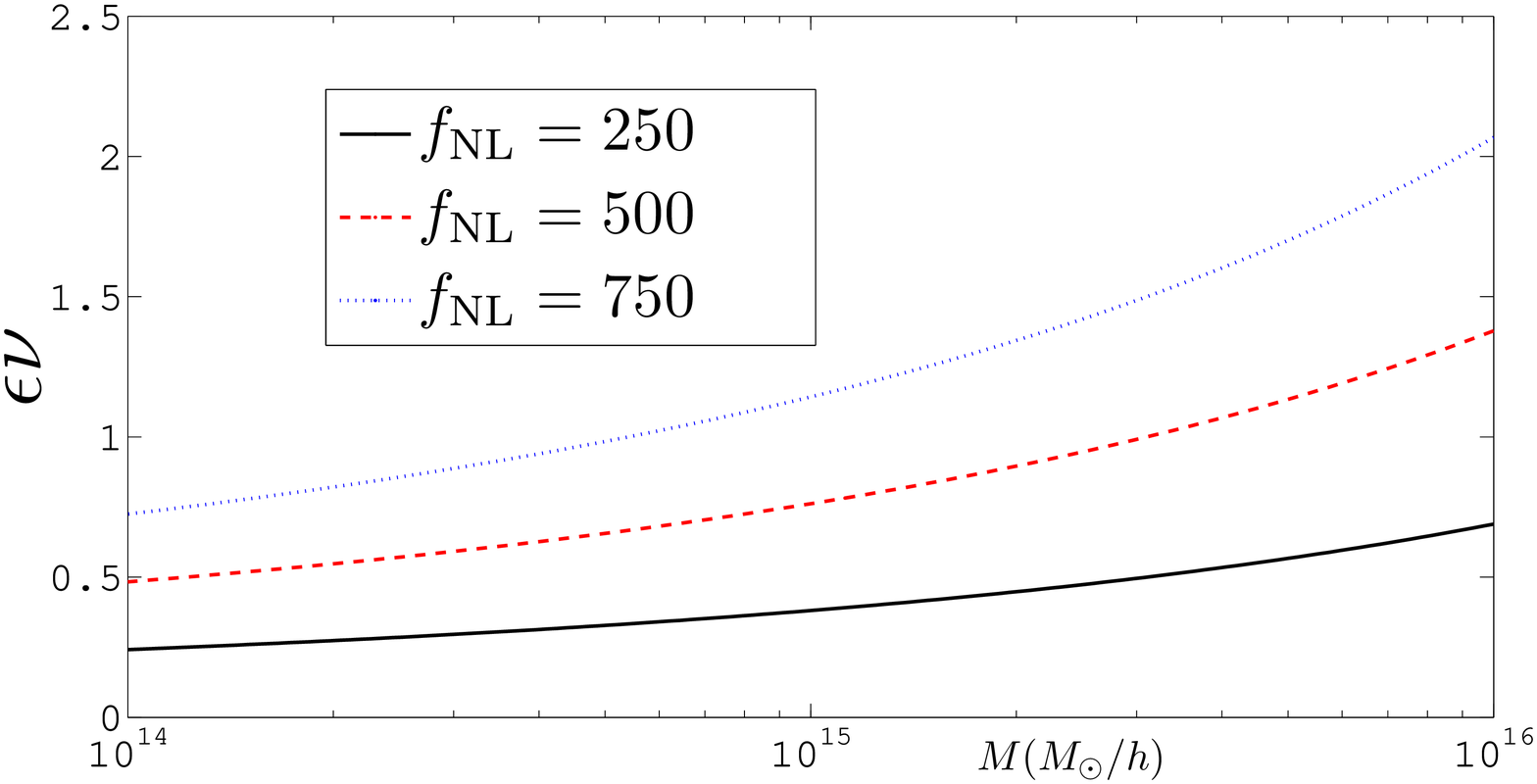}
\caption{\small The quantity $\eps_1\nu$, where $\eps_1$ and $\nu$
  were defined in \eqns{eps1-def} and \eqref{nu-def} respectively. The 
  curves are for redshift $z=1.5$ and three different values of \fnl,
  all probed by the W11 likelihood analysis.} 
\la{fig-epnu}
\end{figure}
with $\eps_1\simeq3\times10^{-4}\fnl$ for a WMAP7
cosmology, and we identify $\eps_1$ with the small parameter \ep\
mentioned above. (Throughout this paper we will use the local model of
NG to compute $\eps_1$, see e.g. D11 for details.) We see
that for $\fnl=750$ at this redshift, $\ep\nu\sim1$ at
$M\sim5\times10^{14}h^{-1}\Msun$, which is well inside the region
explored by W11.

In contrast with the theoretical approaches, the D08 prescription 
constructs a non-Gaussian mass function by convolving a
Gaussian mass function with a Gaussian \emph{probability} distribution
whose mean and variance depend on \fnl. As it turns out, this
prescription results in the only viable mass function currently on the
market which remains stable at large $(\fnl,M,z)$. Since the W11
analysis uses this prescription\footnote{Presumably W11 used the exact
numbers which D08 quote for the parameters of their fit; this detail
is not mentioned in W11.}, it at least does not
suffer from the breakdown which the MVJ, LoVerde \etal\ (LMSV)
\cite{LoVerde:2007ri} or LoVerde \& Smith (LS) \cite{LoVerde:2011iz} 
mass functions would encounter. There is however some cause for
concern. Firstly, the actual parameter values D08 quote are a
numerical fit to simulations which had considerably large Poisson
errors and scatter at large $M$ and $z$ (see Fig.6 of D08). Subsequent
simulations by Pillepich \etal\ (PPH) \cite{Pillepich:2008ka} indicate
that this specific fit does not work very well at the low mass end
either. The main problem however is that this fit applies in a mass
and redshift regime where $|\ep\nu|$ is at most $\sim0.5$, and there
is no reason to expect it to work accurately when $\ep\nu>1$. One
would therefore like to have some analytical understanding of this
extreme tail of the mass function.  

\subsection{A New Stable Non-Gaussian Mass Function}
\la{resum}
\noindent
In this work we present a new analytical non-Gaussian mass function
which is stable in the $|\ep\nu|>1$ regime. As we show in the
Appendix, under some technical assumptions regarding 
the form of primordial non-Gaussianity, which are inspired by a known
perturbative hierarchy in the local model, the perturbative series
appearing in the analysis can be resummed. The excursion set formalism 
then gives a mass function which is valid at \emph{arbitrarily large}
$\nu$ for a given positive \fnl\ (which can also be large, see
below). The formal excursion set result needs to be
modified in order to predict the correct \emph{Gaussian} mass function
when $\fnl\to0$.  Assuming that the Gaussian mass function is well 
described by the Tinker \etal\ form \cite{Tinker:2008ff}, the
``resummed'' mass function we prescribe is  
\be
\left.\frac{dn}{dM}\right|_{\rm Resum} =
\left.\frac{dn}{dM}\right|_{\rm Tinker}\times \Cal{R}_{\rm Resum}\,, 
\la{ngmf-final}
\ee
where the ratio $\Cal{R}_{\rm Resum}$ is given by
\be
\Cal{R}_{\rm Resum}(M,z,\fnl) = \left(1+\eps_1\nu\right)^{-1/2}\exp 
\left[ \frac12\nu^2 + \frac1{\eps_1^2}\left(\eps_1\nu -
  (1+\eps_1\nu)\ln(1+\eps_1\nu)\right) \right]\,. 
\la{resumratio}
\ee
The expression is formally valid for $(1+\eps_1\nu)>0$, and receives a
relative correction of order $\Cal{O}(\eps_1^2(1+\eps_1\nu)^{-1})$ due
to a saddle point approximation (see the Appendix for details). This
shows that although \fnl\ cannot be made arbitrarily large for a given
$\nu$, values of 
the order $\fnl\lesssim1000$ can be easily accomodated over the full
$(M,z)$ range, and the saddle point approximation becomes increasingly 
accurate for large $(M,z)$.  Finally, since the Tinker
\etal\ mass function falls off like $\sim e^{-0.837\delc^2/2\sig^2}$, we
need to redefine the $\nu$ that appears in $\Cal{R}_{\rm Resum}$ by
replacing 
\be
\delc\to\sqrt{q_{\rm Tinker}}\,\delc\,,
\la{delc-replace}
\ee
in \eqn{nu-def} where
$q_{\rm Tinker} = 0.837$. This ensures that the resulting mass
function is still stable for arbitrarily large values of $\nu$. This
correction is similar to the one first introduced by Grossi
\etal\ \cite{Grossi:2007ry} and can be motivated by appealing to a
stochasticity in the spherical collapse barrier
\cite{Maggiore:2009rw}. We emphasize that this modification, 
and the specific value $q_{\rm Tinker}$, is forced on us by our choice
of the Gaussian mass function -- we have no free parameters to play
with in our derivation. 

\fig{fig-pph} shows the non-Gaussian ratios according
to various prescriptions, for values of \fnl\ tested in the PPH
simulations. For better comparison with PPH, these plots assume the
same cosmology that PPH used in their simulations (the WMAP5 values in
their Table 2). We see that in the range plotted, the ratio
$\Cal{R}_{\rm Resum}$ is very close to the LMSV ratio $\Cal{R}_{\rm
  LMSV}$ (which is plotted using a $\nu$ corrected with
$q=0.79$)\footnote{The MVJ and LMSV ratios in our language are
  respectively given by $\Cal{R}_{\rm MVJ} =
  \left(1-\frac13\eps_1\nu\right)^{-1/2} e^{\eps_1\nu^3/6} 
\left(1-\frac{1}{2}\eps_1\nu\left(1-\frac{2}{3}\dot\eps_1
\right)\right)$ and $\Cal{R}_{\rm
  LMSV} = 1 + \frac{1}{6}\eps_1\nu^3\left(
  1-\frac{1}{\nu^2}\left(3-2\dot\eps_1 \right) -
  \frac{2}{\nu^4}\dot\eps_1 \right)$, where $\dot\eps_1\equiv
  d\ln\eps_1/d\ln\sig^2$.}.  
This can be understood analytically since in the regime
$\ep\nu^3<1$, $\Cal{R}_{\rm Resum}$ and $\Cal{R}_{\rm LMSV}$ become 
formally identical when truncated at order $\Cal{O}(\ep\nu)$. The
residual difference is due partly to higher order terms and partly to
the different values of the $q$-correction. \fig{fig-pph} can be
directly compared with  Fig. 5 of PPH, who showed that $\Cal{R}_{\rm
  LMSV}$ with $q=0.79$ performs very well compared to simulations in
this range. Consequently, so does our resummed ratio. 
\begin{figure}[t]
\centering
\includegraphics[width=1.0\textwidth]{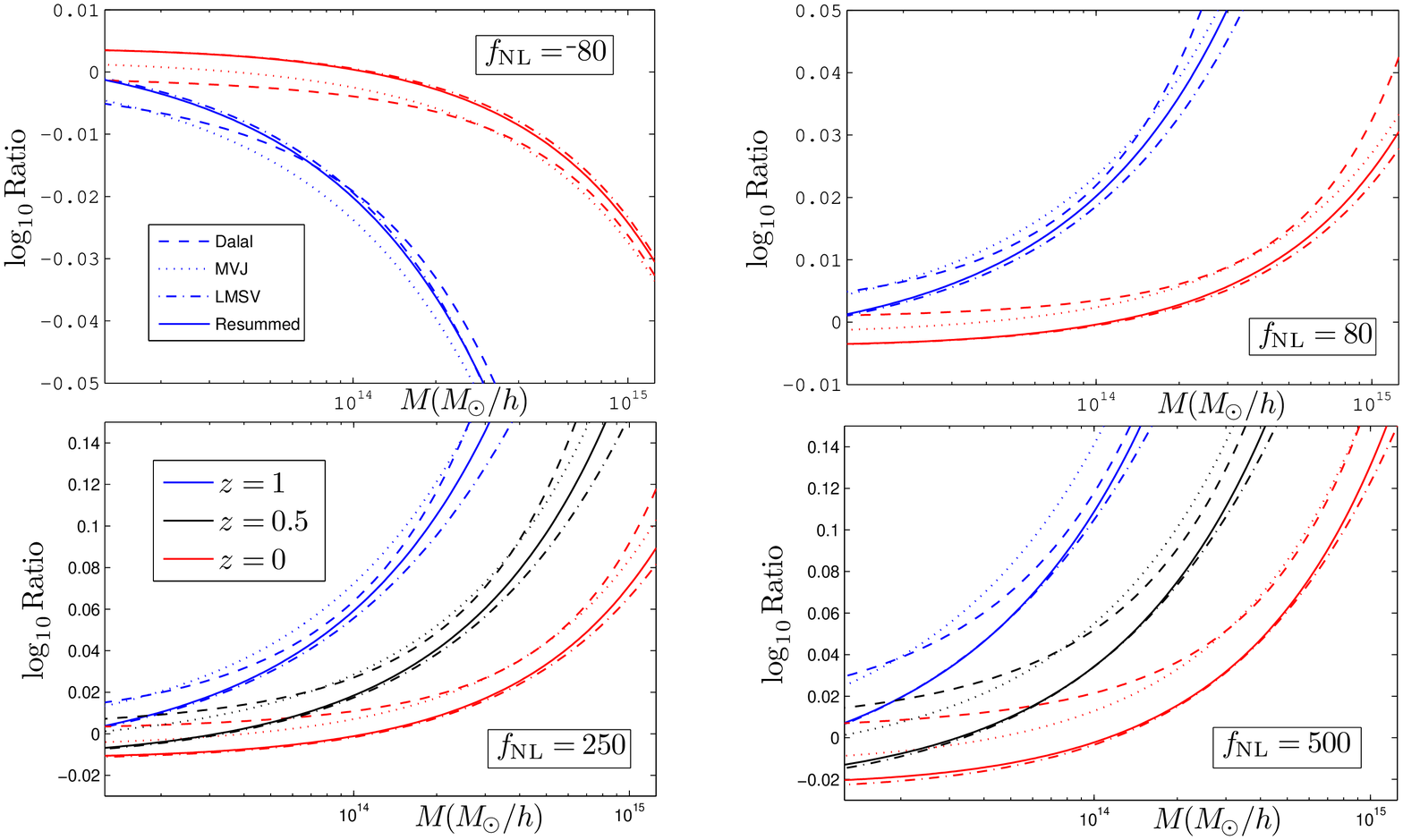}
\caption{\small Non-Gaussian ratios $(dn/dM)|_{\rm NG}/(dn/dM)|_{\rm
    Gauss}$ as per different prescriptions, for values of \fnl\ tested
  in the PPH $N$-body simulations \cite{Pillepich:2008ka}. These plots
  use the same cosmology as used by PPH in their simulations (the
  WMAP5 values in Table 2 of PPH). The curves correspond to resummed
  with $q=q_{\rm Tinker}=0.837$ (solid), LMSV \cite{LoVerde:2007ri}
  with $q=0.79$ (dot-dashed), MVJ \cite{Matarrese:2000iz} with
  $q=0.79$ (dotted), and D08 \cite{Dalal:2007cu} (dashed) calculated
  using the Tinker \etal\ \cite{Tinker:2008ff} Gaussian mass
  function. The different sets of curves in each panel correspond to
  different redshifts. The lower panels show, from left to right,
  redshifts $z=1$, $0.5$ and $0$. The upper panels show $z=1$ and $0$,
  with $z=1$ corresponding to the lower set of curves in the left
  panel ($\fnl=-80$) and vice-versa in the right panel
  ($\fnl=+80$). We see that in this regime of parameter space, the
  resummed prescription (which has no free parameters once the
  Gaussian mass function is chosen) lies very close to LMSV with
  $q=0.79$. These plots can be directly compared with Fig. 5 of PPH,
  who showed that LMSV with $q=0.79$ fares very well compared to 
  simulations. Consequently, so does the resummed prescription.}  
\la{fig-pph}
\end{figure}

For large negative \fnl\ (or large $\nu$) such that $\ep\nu<-1$, our
prescription breaks down. Nevertheless, there is a very simple way of
extending the mass function to this regime. We notice that for
$|\ep\nu|\ll1$, it is approximately true that $\Cal{R}_{\rm
  Resum}(M,z,-\fnl) \simeq \Cal{R}_{\rm Resum}(M,z,\fnl)^{-1}$. We
therefore \emph{define} the resummed ratio for negative \fnl\ and
arbitrary $\nu$ by 
\be
\Cal{R}_{\rm Resum}(M,z,-\fnl) \equiv \left( \Cal{R}_{\rm
  Resum}(M,z,\fnl) \right)^{-1} ~~;~~ \fnl>0\,.
\la{resumratio-negfnl}
\ee
Of course this is completely \emph{ad hoc}, but we see from
\fig{fig-pph} that the prescription compares extremely well with
$\Cal{R}_{\rm LMSV}$ at values of $\fnl<0$ that were tested in the PPH 
simulations. It will be very interesting to see how the resummed ratio
compares with simulations at high masses and redshifts, for both
positive and negative \fnl\ values.

\section{Constraints on \fnl}
\la{fisher}
\noindent
\fig{fig-pph} also shows that for a given positive \fnl, the D08
ratio $\Cal{R}_{\rm Dalal}$ is consistently higher than $\Cal{R}_{\rm 
  Resum}$ and therefore predicts more high mass halos. This trend
remains true even in the regime $\ep\nu>1$, as \fig{fig-ratios}
shows. It is then worth asking how sensitive the analysis of clusters
\emph{a la} W11 is to this difference\footnote{We
  note that in contrast to W11, earlier analyses
  \cite{Hoyle:2010ce,Enqvist:2010bg} showed a significant tension with
  standard \Lam CDM. It would then seem to be more interesting to
  perform our analysis on the set of clusters used in these analyses
  rather than W11. In fact, these analyses can be shown to have used a
  biased statistic, and removing this bias makes them consistent with
  the analysis of W11 \cite{futurework}. We ultimately choose to study
  W11 because we find it easier to approximate the selection function
  for the Sunyaev-Zeldovich clusters measured by SPT.}.

We use a Fisher analysis to make a ``forecast'' for the kind of data
that SPT has already observed (specifically the subset used by W11).  
We set our fiducial cosmology to the WMAP7 values given earlier, with
$\fnl=0$. We marginalise over $\sig_8$ with a WMAP7 prior, but assume 
perfect knowledge of all other cosmological parameters. This is not
expected to significantly affect our conclusions. We bin in redshift
between $0.3\leq z\leq 2$, with a spacing $\Del z=0.05$. The lower
limit is the same used by W11, and we have checked that the analysis
is insensitive to increasing the upper limit. We assume that the
survey sees all clusters in any given redshift bin above a certain
constant mass threshold $M_{\rm lim}$, which is approximately true for
a Sunyaev-Zeldovich survey such as SPT \cite{Vanderlinde:2010eb}. We
marginalise over $M_{\rm lim}$ with a lognormal prior of $30\%$ in
$M_{\rm lim}$, in keeping with typical mass uncertainties quoted by
W11. The fiducial value for $M_{\rm lim}$ is chosen such that the
total expected number of clusters with masses above $M_{\rm lim}$ and
at redshifts $z\geq0.3$ in the fiducial cosmology, is roughly the same
as in the W11 analysis, which is $26$. For $\fnl=0$, (i.e. -- using
the Tinker \etal\ mass function) this gives us a fiducial value
$M_{\rm lim} =  8.5\times10^{14}h^{-1}\Msun$ for $f_{\rm sky}=0.06$.
\begin{figure}[t]
\centering
\includegraphics[height=0.3\textheight]{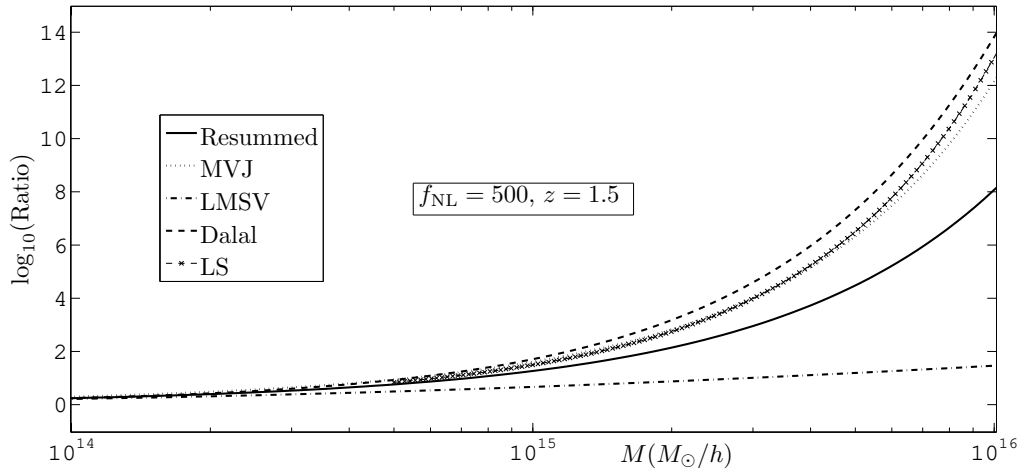}
\caption{\small Non-Gaussian ratios $(dn/dM)|_{\rm NG}/(dn/dM)|_{\rm
    Gauss}$ as per different prescriptions, for $\fnl=500$ and
  $z=1.5$. Comparing with \fig{fig-epnu} we see that the curves now
  enter the regime $\ep\nu>1$. The curves correspond to resummed with
  $q=q_{\rm Tinker}=0.837$ (solid), LMSV (dot-dashed) and MVJ (dotted)
  with $q=0.79$, D08 (dashed) with the Tinker \etal\ Gaussian, and the
  ``log-Edgeworth'' ratio prescribed by LS  \cite{LoVerde:2011iz}
  (cross-dashed). The D08 prescription continues to systematically
  predict a larger number of massive halos than the resummed.}
\la{fig-ratios}
\end{figure}

We then assume a joint likelihood given by a product of independent
Poisson probabilities for each redshift bin\footnote{This is similar
  to what is done by W11, who use a product of independent Poisson
  probabilities over a number of bins in the space of redshift and
  detection significance which is roughly their mass proxy. 
  As can be seen from the methods of \Cite{Hu:2002we},
  sample covariance should be negligible for this survey.}, and construct the Fisher
matrix for parameters $\tht_a=(M_{\rm lim},\sig_8,\fnl)$
\cite{Holder:2001db} 
\be
F_{ab} = \sum_{i=1}^{n_{\rm bins}}\frac1{\mu_i}
\frac{\p\mu_i}{\p\tht_a} \frac{\p\mu_i}{\p\tht_b}\,,
\la{fishermatrix}
\ee
where $\mu_i$ is the expected number of clusters in the $i^{\rm th}$
redshift bin, evaluated as the integral of $(dn/dM)(dV/dz)$ over this
bin, for masses above $M_{\rm lim}$. The results are in
\tab{tab-fisher-spt}, which shows the marginalised errors on \fnl.  
\begin{table}[t]
\begin{tabular}{|c|c|c|}\hline
\multicolumn{3}{|c|}{Marginal $\sig_{\fnl}$, W11 equivalent} \\ [0.5ex]\hline
 \ph{hi} Prescription  \ph{hi} & $\ph{00}\fnl=0\ph{00}$ & $\ph{0}\fnl=500\ph{0}$ \\ [1ex]\hline \hline
 D08 & $298$ & $463$ \\ [0.5ex]\hline
 Resummed & $457$ & $599$ \\ [0.5ex]\hline
 MVJ & $487$ & $288$ \\ [0.5ex]\hline
 LMSV & $498$ & $1366$ \\ [0.5ex]\hline
\end{tabular}
\caption{\small The marginal Fisher error $\sig_{\fnl}$ as per
  different prescriptions, for two different choices of fiducial \fnl,
  for a survey that is approximately equivalent to the subset of the
  SPT clusters analysed in W11 (see text for details). The
  marginalisation is over $\sig_8$ with a WMAP7 prior and the
  threshold mass $M_{\rm lim}$ with a $30\%$ lognormal prior (see
  text). We see that the resummed and D08 prescriptions predict
  comparable errors in all cases, while the MVJ and LMSV predict
  significantly different values. See text for a discussion.}
\la{tab-fisher-spt}
\end{table}
We see that the D08 mass function results in a marginal error
$\sig_{\fnl}\simeq300$, which can be compared with the error quoted by
W11 which is $450$. This shows that our analysis works; the fact that
our marginal error for D08 is smaller, is consistent with our
simplifying assumptions regarding the cosmological parameters and
selection function. The errors predicted by the resummed mass function
are comparable to that predicted by the D08 prescription, indicating
that using either prescription would lead to similar results. Since
the errors are large however, one might wonder if the situation would
change if the fiducial \fnl\ were some large value consistent with the
error. \tab{tab-fisher-spt} shows that the predicted errors from the
D08 and resummed mass functions are still comparable at a fiducial
$\fnl=500$ (in this case we used a fiducial $M_{\rm
  lim}=1.05\times10^{15}h^{-1}\Msun$ to get a total expected number
close to $26$).  
\begin{table}[t]
\begin{tabular}{|c|c|c|c|c|}\hline
\multicolumn{5}{|c|}{$\sig_{\fnl}$, \emph{Planck}-like} \\ [0.5ex]\hline
&\multicolumn{2}{|c|}{\bf Conditional}&\multicolumn{2}{|c|}{\bf Marginal} \\ [0.5ex]\hline
 \ph{hi} Prescription  \ph{hi} & $\ph{00}\fnl=0\ph{00}$ &
 $\ph{0}\fnl=500\ph{0}$ & $\ph{00}\fnl=0\ph{00}$ &
 $\ph{0}\fnl=500\ph{0}$ \\ [1ex]\hline \hline
 D08 & $9.1$ & $9.2$ & $45.7$ & $77.9$ \\ [0.5ex]\hline
 Resummed & $11.0$ &$11.0$ & $64.0$ & $80.5$ \\ [0.5ex]\hline
 MVJ & $10.8$ & $6.6$ & $72.4$ & $41.7$ \\ [0.5ex]\hline
 LMSV & $12.0$ & $16.1$ & $69.6$ & $310$ \\ [0.5ex]\hline
\end{tabular}
\caption{\small Fisher errors on \fnl\ for a final  \emph{Planck}-like
  survey, with a $2\%$ prior on $\sig_8$ and a fiducial threshold
  $M_{\rm lim}=5\times10^{14}h^{-1}\Msun$ with a $10\%$ lognormal
  prior and $f_{\rm sky}=0.8$. The values under ``conditional'' assume perfect knowledge of
  all parameters except \fnl, and represent what is possible if
  parameter degeneracies are broken using measurements of say, the
  clustering of clusters
  \cite{Cunha:2010zz,Sartoris:2010cr,Fedeli:2010ud}.}  
\la{tab-fisher-planck}
\end{table}

\tab{tab-fisher-planck} shows the results of a
Fisher forecast for a final \emph{Planck}-like SZ survey
\cite{Geisbuesch:2006fr}, assuming a $2\%$ prior on $\sig_8$, a
threshold $M_{\rm lim}=5\times10^{14}h^{-1}\Msun$ with a $10\%$ prior,
and binning in redshift between $0.1\leq z\leq 2$ with a spacing $\Del
z=0.05$ and $f_{\rm sky}=0.8$. In this case we also display the conditional error (i.e. --
assuming perfect knowledge of other parameters), as being
representative of what can be achieved when parameter degeneracies are
broken using measurements of say, the clustering of clusters 
\cite{Cunha:2010zz,Sartoris:2010cr,Fedeli:2010ud}. We see that the
predicted conditional and marginal errors for \fnl\ are again
comparable for the two mass functions\footnote{This might seem
  surprising given the difference between the magnitudes of these mass
  functions. In fact, it might appear from \fig{fig-ratios} that the
  MVJ prescription should be closer to D08. Note however, that the
  Fisher matrix \eqref{fishermatrix} depends not only on the
  integrated mass function, but also on its derivative with respect to
  \fnl, leading to a complicated interplay which is difficult to
  predict simply by examining plots of the mass function or
  non-Gaussian ratio.}. 

It appears therefore, that using the D08 fit at
$\ep\nu>1$, as opposed to a theoretically motivated mass function
such as the resummed, does not introduce a significant effect in the
error predicted for \fnl\ from number counts of clusters. However,
since the actual number of halos predicted at any given \fnl\ is
different for these two mass functions, we must also worry about a
possible bias in the central value of \fnl. Estimating this properly
would require a full-fledged Monte Carlo analysis that accurately
accounts for survey selection functions, which is work in
progress. For now, we perform a cruder analysis. For parameters
corresponding to the final \emph{Planck}-like survey mentioned above, we
assume a fiducial cosmology of the WMAP7 \Lam CDM, but this time with
a non-zero value of $\fnl={\fnl}_\ast$ using the D08 prescription. We
then ``analyse data'' using the resummed prescription, by constructing
an approximate likelihood given by 
\be
\Cal{L}(\fnl,M_{\rm lim}) \propto 
 \exp\left[-\frac12  \sum_{i=1}^{n_{\rm bins}}
   \left(\frac{\left(\mu_i^{\rm Resum}- \mu_{i\ast}^{\rm
       Dalal}\right)^2}{\mu_i^{\rm Resum}} + \ln\mu_i^{\rm
     Resum}\right) \right]\,, 
\la{bias-likelihood}
\ee
where $\mu_{i\ast}^{\rm Dalal} = \mu_i^{\rm Dalal}({\fnl}_\ast,M_{{\rm
    lim}\ast})$ is the fiducial expectation value in the $i^{\rm th}$
bin, which we treat as ``data'', and $\mu_i^{\rm Resum} = \mu_i^{\rm
  Resum}(\fnl,M_{\rm lim})$ is the model expectation value computed
using the resummed prescription at the $(\fnl,M_{\rm lim})$ values
being analysed. (We keep $\sig_8$ fixed in this exercise.) We
marginalise this likelihood over $M_{\rm lim}$ with a $10\%$ lognormal
prior (assuming a flat prior on \fnl), to obtain a posterior
probability distribution for \fnl. \fig{fig-bias} shows the results
for ${\fnl}_\ast=100$ and $500$. 
We see that for ${\fnl}_\ast=100$, the resummed posterior
$p(\fnl|{\fnl}_\ast=100)$ peaks close to ${\fnl}_\ast$ and there is no
statistically significant bias; ${\fnl}_\ast$ lies less than one
standard deviation away from the peak value. For ${\fnl}_\ast=500$ on
the other hand, the value ${\fnl}_\ast$ lies far in the tail (more
than $4$ standard deviations away from the peak) of the distribution
$p(\fnl|{\fnl}_\ast=500)$, indicating a statistically significant
bias. A similar calculation for the SPT
equivalent parameter values used in  \tab{tab-fisher-spt}
shows that with the current data there would be no detectable
difference between analyses based on the resummed or D08
prescriptions. We can conclude that systematic effects
from the choice of non-Gaussian mass function could become important
as the data improves, if non-Gaussianity is scale dependent and
results in a large \fnl\ on cluster scales. In this case it would be
necessary to properly calibrate the non-Gaussian mass function in its
extreme tail.
\begin{figure}[t]
\centering
\includegraphics[height=0.25\textheight]{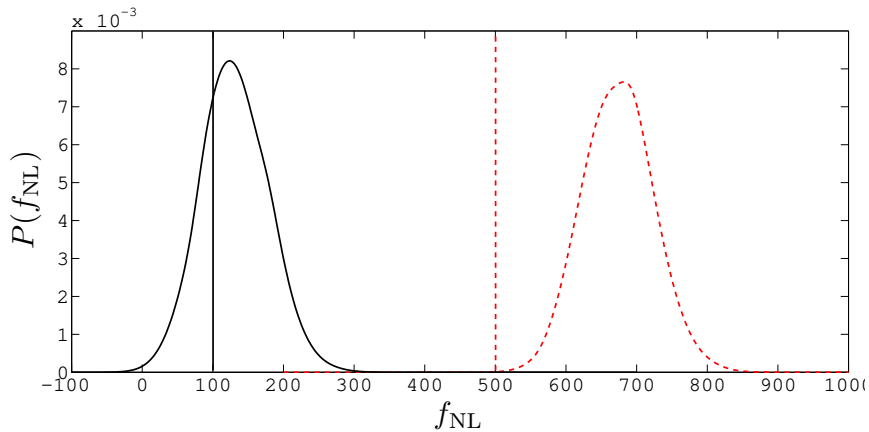}
\caption{\small Testing for bias between D08 and resummed for a final {\em Planck} like survey : Posterior
  probability distribution $p(\fnl|{\fnl}_\ast)$ computed by
  marginalising the approximate likelihood \eqref{bias-likelihood} for
  the D08 ``data'' given the resummed ``model'', over $M_{\rm lim}$
  with a $10\%$ lognormal prior. The curves are for ${\fnl}_\ast=100$
  (solid,black) and ${\fnl}_\ast=500$ (dashed,red). The corresponding
  vertical lines mark the value of the respective ${\fnl}_\ast$. We
  see a statistically significant relative bias between the resummed
  and D08 prescriptions for large ${\fnl}_\ast$, while at smaller 
  ${\fnl}_\ast$ the prescriptions are statistically consistent with
  each other.}  
\la{fig-bias}
\end{figure}

\section{Summary \& Conclusions}
\la{conclude}
\noindent
Number counts of clusters offer a very interesting probe of the
extreme tail of the mass function of collapsed objects, and thereby
the statistics of the primordial curvature fluctuations. In this work
we developed a new prescription for calculating the modification of
the mass function in the presence of primordial non-Gaussianity
(\eqns{ngmf-final} and \eqref{resumratio}). The key features of
this ``resummed'' prescription are that it is theoretically well
motivated (as opposed to a fit to simulations like D08
\cite{Dalal:2007cu}) as well as stable in the extreme region
$|\ep\nu|>1$ (unlike all the other currently available theoretical
prescriptions \cite{Matarrese:2000iz, LoVerde:2007ri, Maggiore:2009rx, 
  D'Amico:2010ta, LoVerde:2011iz}).  We also showed that this resummed
prescription compares very well against the results of $N$-body
simulations in the literature (see \fig{fig-pph}), despite having no
free parameters left once its Gaussian limit is fixed as the Tinker
\etal\ \cite{Tinker:2008ff} mass function.   

While this might appear to be an academic issue, we showed that the
current quality of data forces any likelihood analysis to enter the
regime $|\ep\nu|>1$. It is then important to analyse how far one can
trust the chosen prescription for the non-Gaussian mass function.
The specific example we dealt with was the W11
\cite{Williamson:2011jz} analysis of a subset of the current SPT
clusters, which used the D08 prescription to compute the likelihood.  
We used Fisher matrix techniques to analyse the consequences of
changing the prescription from D08 to resummed, and found that for the
current data, both prescriptions are expected to give approximately identical
results. As the data improve however, there is a possibility of
introducing a statistically significant bias in the analysis, due to
the choice of prescription. It will then become important to evaluate
which is the better method, which will require improved $N$-body
simulations. This would be especially true if non-Gaussianity is scale
dependent and results in a large value of \fnl\ on cluster scales
\cite{Riotto:2010nh,Byrnes:2009pe}. One would then need to calibrate
the mass function using $N$-body simulations which may have to be
specially tailored (with larger particle size, say) to probe clusters
with high masses, at large redshifts and in the presence of large \fnl.

\acknowledgments
\noindent
We thank L. Sriramkumar and other organisers of the
Primordial Features and Non-Gaussianities (PFNG) meeting at the
Harish-Chandra Research Institute in December 2010 where this work was
initiated. We also thank Ravi Sheth for helpful discussions. CG is
funded by the Beecroft Institute for Particle Astrophysics and
Cosmology. SH is supported by the Academy of Finland grant 131454.

\appendix
\section{The Resummed Mass Function}
\noindent
In this Appendix we sketch the derivation of the resummed mass
function presented in the text, describing all the approximations that
enter. The non-Gaussian halo mass function derived from the
excursion set formalism for spherical collapse is
\be
\left.\frac{dn}{dM}\right|_{\rm NG,exc} = \frac{\bar\rho}{M^2}f_{\rm 
  NG,exc}\left| \frac{d\ln\sig}{d\ln M} \right|\,,
\la{ngmf-es}
\ee
where the multiplicity $f_{\rm NG,exc}$ for large masses is given by
(see D11 \cite{D'Amico:2010ta})
\be
f_{\rm NG,exc}(\nu,\eps_1,\eps_2,\ldots) = \sqrt{\frac2\pi} \,\nu\,
\exp\left[\sum_{n=3}^\infty \frac{(-1)^n}{n!} \eps_{n-2}\p_\nu^n
  \right] \bigg\{ e^{-\frac12\nu^2} + \ldots \bigg\}\,,
\la{ngmult-es}
\ee
where $\nu(M,z)$ was defined in \eqn{nu-def} and
$\eps_{n-2}\equiv\avg{\hat\del_M^n}/\sig_M^n$ are the normalised connected 
moments of the linearly extrapolated, smoothed density field. This
expression assumes that the functions $\eps_j$ are all constant with
scale in the regime of interest, which is reasonable at least in the
local model (see e.g. Fig.1 of D11). If \fnl\ is the only NG parameter
allowed, then these are perturbatively ordered : $\eps_j \sim (\fnl
\sqrt A)^j$, with $A$ the normalisation of the power spectrum. The
ellipsis in \eqn{ngmult-es} indicates terms arising from
``unequal-time'' correlations as discussed in
\Cites{Maggiore:2009rx,D'Amico:2010ta}. In the large mass (or large
$\nu$) limit, these will be of the form $\sim
e^{-\frac12\nu^2}\Cal{O}(\ep\nu)$. Experience with the calculation of 
D11 indicates that the action of the exponential derivative on all
these terms will result in a \emph{single} common exponential
prefactor multiplying a series expansion in $\ep\nu$. When
$|\ep\nu|\sim1$, the polynomial-type terms should resum, but it is hard
(if not impossible) to predict the resummed form. We will therefore
ignore these terms and concentrate on the exponential, with the
\emph{post hoc} justification that the approximations appear to work
very well at intermediate masses and redshifts where non-Gaussian
$N$-body simulations have been performed. 

We are then after the quantity
\begin{align}
f_{\rm NG,exc}(\nu,\eps_1,\eps_2,\ldots) &\approx \sqrt{\frac2\pi}
\,\nu\, \exp\left[ \sum_{n=3}^\infty \frac{(-1)^n}{n!}
  \eps_{n-2}\p_\nu^n \right]  e^{-\frac12\nu^2}\nonumber\\
&=\sqrt{\frac2\pi} \,\nu\, \lamint{e^{\phi(\lam)}}\,,
\la{ngmult-es-fourier}
\end{align}
where we defined
\be
\phi(\lam) \equiv i\lam\nu +
\sum_{n=2}^{\infty}\frac{(-i\lam)^n}{n!}\eps_{n-2} ~~;~~
\eps_0\equiv1\,. 
\la{phi-Q-def}
\ee
The integral in \eqn{ngmult-es-fourier} can in principle be done using
a saddle point approximation with calculable
corrections, e.g. along the lines presented in D11. Unfortunately, as
D11 showed, a perturbative treatment will necessarily require the
condition $|\ep\nu|<1$, which is not sufficient for our purpose. Going
beyond this technical barrier requires knowledge of an infinite
number of connected moments of the density field.  A key
simplification occurs if we assume (inspired by the heirarchy
$\eps_j\sim(\fnl\sqrt A)^j$) the \emph{exact} relations 
\be
\eps_j = \ep^j ~~;~~ \ep \equiv \eps_1\,.
\la{epsj-assume}
\ee
One could treat this as defining a specific model of
non-Gaussianity. Throughout, we will numerically compute $\eps_1$ as
in the local model of NG. The series in $\phi(\lam)$ can then be
re-arranged to bring the function into closed form, 
\be
\phi(\lam) = i\lam\nu + \frac1{\ep^2} \left( e^{-i\lam\ep} + i\lam\ep
-1\right)\,, 
\la{phi-closedform}
\ee
with derivatives
\be
\phi^\prime(\lam) = i\left(\nu +
\frac1\ep\left(1-e^{-i\lam\ep}\right)\right) ~~;~~
\phi^{\prime\prime}(\lam) = -e^{-i\lam\ep}\,.
\la{phi-derivs}
\ee
A saddle point now exists at $\lam=\lam_\ast$ where
$\phi^\prime(\lam_\ast)=0$, i.e. 
\be
e^{-i\lam_\ast\ep} = 1+\ep\nu\,,
\la{lam-saddlept}
\ee
provided we have $\phi^{\prime\prime}(\lam_\ast)=-(1+\ep\nu)<0$,
i.e. if $\ep\nu>-1$. The leading saddle point result for the
multiplicity follows from setting $\int d\lam/\sqrt{2\pi}
e^{\phi(\lam)}=
e^{\phi(\lam_\ast)}(|\phi^{\prime\prime}(\lam_\ast)|)^{-1/2}$, which
gives
\begin{align}
f_{\rm NG,exc}(\nu,\ep) &\approx \sqrt{\frac2\pi}\, \nu\,
\left(1+\ep\nu\right)^{-1/2} \exp\left[\frac1{\ep^2}\left(\ep\nu - 
  (1+\ep\nu)\ln(1+\ep\nu)\right)\right] \nonumber\\
&\equiv f_{\rm NG,app}(\nu,\ep)\,,
\la{ngmult-spapp}
\end{align}
the subscript ``app'' reminding us of the approximations involved in
the derivation.

One can also estimate the error involved in the saddle point
approximation. The calculation proceeds along the lines discussed in
Appendix D of D11, by first using the series representation of
$\phi(\lam)$ in the integral of $e^{\phi(\lam)}$, and then Taylor
expanding to get (after a change of variables) the exact result
\be
\lamint{e^{\phi(\lam)}} = e^{\phi(\lam_\ast)}(1+\ep\nu)^{-1/2}
\int_{-\infty}^{\infty}\frac{dy}{\sqrt{2\pi}}e^{-\frac12y^2}\left[ 1+
  \frac{(-iy)^3}{3!} \tau^{1/2} + \frac{(-iy)^4}{4!}\tau + \frac1{2!}
  \left(\frac{(-iy)^3}{3!}\right)^2 \tau + \ldots \right]\,,
\la{saddlept-corr}
\ee
where $\tau \equiv \ep^2/(1+\ep\nu) > 0$. Clearly, terms involving
half-integer powers of $\tau$ will not contribute, and hence the
relative correction to 
\eqn{ngmult-spapp} is of order $\Cal{O}(\tau)$ and is calculable in
principle to arbitrary order in $\tau$. Since $\ep\sim
3\times10^{-4}\fnl$, the approximation remains valid even for
considerably large positive values of \fnl, and becomes increasingly
better for large $\nu$. Negative \fnl\ values do not fare
as well though, due to the restriction $\ep\nu>-1$. See the main text
however for a simple way of extending the mass function to $\fnl<0$,
which appears to work well in the regime where $N$-body simulations
have been performed.

It is straightforward to check that for fixed \fnl, the multiplicity
$f_{\rm NG,app}(\nu,\ep)$ is monotonically decreasing with $\nu$ at
large $\nu$, and the mass function is therefore stable in its extreme
tail, as needed. One can also easily check (using
$\ln(1+x)=-\sum_{n=1}^\infty(-x)^n/n$) that  the limit $\ep\to0$
recovers the Gaussian Press-Schechter result,  
\be
\lim_{\ep\to0} f_{\rm NG,app}(\nu,\ep) = \sqrt{\frac2\pi}\,\nu\,
e^{-\frac12\nu^2} \equiv f_{\rm PS}(\nu)\,.
\la{ngmult-gausslim}
\ee
In fact, retaining the leading terms in \ep\ recovers the MVJ
exponential prefactor 
\be
f_{\rm NG,app}(\nu,\ep) \to \sqrt{\frac2\pi}\,\nu\,
e^{-\frac12\nu^2(1-\frac13\ep\nu +\ldots)}
\left(1+\Cal{O}(\ep\nu)\right) \,.
\la{ngmult-MVJlim}
\ee
%
Since the Press-Schechter result is known to perform
badly when compared with $N$-body simulations, we will follow the
usual practice in the literature and prescribe a \emph{ratio} of
non-Gaussian and Gaussian mass functions. Assuming that the Gaussian
mass function is well described by the Tinker \etal\ form
\cite{Tinker:2008ff}, the final mass function to use is given by
\eqn{ngmf-final}, with a modified definition of $\nu$ as discussed in
the main text.


\begin{thebibliography}{10}


\bibitem{Maldacena:2002vr}
  J.~M.~Maldacena,
  JHEP {\bf 0305}, 013 (2003)
  [arXiv:astro-ph/0210603].

\bibitem{Acquaviva:2002ud}
  V.~Acquaviva, N.~Bartolo, S.~Matarrese and A.~Riotto,
  Nucl.\ Phys.\  B {\bf 667}, 119 (2003)
  [arXiv:astro-ph/0209156].

\bibitem{Desjacques:2010jw}
  V.~Desjacques and U.~Seljak,
  Class.\ Quant.\ Grav.\  {\bf 27}, 124011 (2010)
  [arXiv:1003.5020 [astro-ph.CO]].

\bibitem{Komatsu:2010fb}
  E.~Komatsu {\it et al.} [ WMAP Collaboration ],
  Astrophys.\ J.\ Suppl.\  {\bf 192}, 18 (2011).
  [arXiv:1001.4538 [astro-ph.CO]].

\bibitem{Slosar:2008hx}
  A.~Slosar, \etal,
  JCAP {\bf 0808} (2008) 031
  [arXiv:0805.3580 [astro-ph]].

\bibitem{Sefusatti:2009xu}
  E.~Sefusatti, \etal,
  JCAP {\bf 0912}, 022 (2009).
  [arXiv:0906.0232 [astro-ph.CO]].

\bibitem{LoVerde:2007ri}
  M.~LoVerde, A.~Miller, S.~Shandera and L.~Verde,
  JCAP {\bf 0804} (2008) 014
  [arXiv:0711.4126 [astro-ph]].

\bibitem{Riotto:2010nh}
  A.~Riotto, M.~S.~Sloth,
  Phys.\ Rev.\  {\bf D83}, 041301 (2011).
  [arXiv:1009.3020 [astro-ph.CO]].

\bibitem{Byrnes:2009pe}
  C.~T.~Byrnes, S.~Nurmi, G.~Tasinato, D.~Wands,
  JCAP {\bf 1002}, 034 (2010).
  [arXiv:0911.2780 [astro-ph.CO]].

\bibitem{Fedeli:2010ud}
  C.~Fedeli, C.~Carbone, L.~Moscardini and A.~Cimatti,
  arXiv:1012.2305 [astro-ph.CO].

\bibitem{Cunha:2010zz}
  C.~Cunha, D.~Huterer and O.~Dore,
  Phys.\ Rev.\  D {\bf 82} (2010) 023004
  [arXiv:1003.2416 [astro-ph.CO]].

\bibitem{Sartoris:2010cr}
  B.~Sartoris, \etal,
  Mon.\ Not.\ Roy.\ Astron.\ Soc.\  {\bf 407}, 2339 (2010),
  arXiv:1003.0841 [astro-ph.CO].

\bibitem{Jimenez:2009us}
  R.~Jimenez, L.~Verde,
  Phys.\ Rev.\  {\bf D80}, 127302 (2009).
  [arXiv:0909.0403 [astro-ph.CO]].

\bibitem{Cayon:2010mq}
  L.~Cay\'on, C.~Gordon, J.~Silk,
    [arXiv:1006.1950 [astro-ph.CO]], MNRAS in press.

\bibitem{Hoyle:2010ce}
  B.~Hoyle, R.~Jimenez, L.~Verde,
  [arXiv:1009.3884 [astro-ph.CO]].

\bibitem{Enqvist:2010bg}
  K.~Enqvist, S.~Hotchkiss, O.~Taanila,
  [arXiv:1012.2732 [astro-ph.CO]], JCAP in press.

\bibitem{Williamson:2011jz}
  R.~Williamson, \etal,
  [arXiv:1101.1290 [astro-ph.CO]].

\bibitem{Dalal:2007cu}
  N.~Dalal, O.~Dore, D.~Huterer and A.~Shirokov,
  Phys.\ Rev.\  D {\bf 77} (2008) 123514
  [arXiv:0710.4560 [astro-ph]].

\bibitem{Matarrese:2000iz}
  S.~Matarrese, L.~Verde and R.~Jimenez,
  Astrophys.\ J.\  {\bf 541} (2000) 10
  [arXiv:astro-ph/0001366].

\bibitem{Maggiore:2009rx}
  M.~Maggiore, A.~Riotto,
  Astrophys.\ J.\  {\bf 717}, 526-541 (2010).
  [arXiv:0903.1251 [astro-ph.CO]].

\bibitem{D'Amico:2010ta}
  G.~D'Amico, M.~Musso, J.~Nore\~na, A.~Paranjape,
  JCAP {\bf 1102}, 001 (2011).
  [arXiv:1005.1203 [astro-ph.CO]].

\bibitem{LoVerde:2011iz}
  M.~LoVerde, K.~M.~Smith,
  [arXiv:1102.1439 [astro-ph.CO]].

\bibitem{Bardeen:1985tr}
  J.~M.~Bardeen, J.~R.~Bond, N.~Kaiser and A.~S.~Szalay,
  Astrophys.\ J.\  {\bf 304}, 15 (1986).

\bibitem{Sugiyama:1994ed}
  N.~Sugiyama,
  Astrophys.\ J.\ Suppl.\  {\bf 100}, 281 (1995)
  [arXiv:astro-ph/9412025].

\bibitem{Press:1973iz}
  W.~H.~Press andP.~Schechter,
  Astrophys.\ J.\  {\bf 187}, 425 (1974).

\bibitem{Bond:1990iw}
  J.~R.~Bond, S.~Cole, G.~Efstathiou and N.~Kaiser,
  Astrophys.\ J.\  {\bf 379}, 440 (1991).

\bibitem{Sheth:1999mn}
  R.~K.~Sheth, G.~Tormen,
  Mon.\ Not.\ Roy.\ Astron.\ Soc.\  {\bf 308}, 119 (1999).
  [astro-ph/9901122].


\bibitem{Pillepich:2008ka}
  A.~Pillepich, C.~Porciani, O.~Hahn,
  Mon.\ Not.\ Royal\ Astron.\ Soc., {\bf 402}, 191-206 (2010).
  [arXiv:0811.4176 [astro-ph]].

\bibitem{Tinker:2008ff}
  J.~L.~Tinker, \etal,
  Astrophys.\ J.\  {\bf 688}, 709-728 (2008).
  [arXiv:0803.2706 [astro-ph]].

\bibitem{Grossi:2007ry}
  M.~Grossi, \etal,
  Mon.\ Not.\ Roy.\ Astron.\ Soc.\  {\bf 382}, 1261 (2007).
  [arXiv:0707.2516 [astro-ph]].

\bibitem{Maggiore:2009rw}
  M.~Maggiore, A.~Riotto,
  Astrophys.\ J.\  {\bf 717}, 515-525 (2010).
  [arXiv:0903.1250 [astro-ph.CO]].

\bibitem{futurework}
S. Hotchkiss, [arXiv:1105.3630 [astro-ph.CO]], JCAP in press. 

\bibitem{Vanderlinde:2010eb}
  K.~Vanderlinde, \etal,
  Astrophys.\ J.\  {\bf 722}, 1180-1196 (2010).
  [arXiv:1003.0003 [astro-ph.CO]].

\bibitem{Hu:2002we}
  W.~Hu, A.~V.~Kravtsov,
  Astrophys.\ J.\  {\bf 584}, 702-715 (2003).
  [astro-ph/0203169].

\bibitem{Holder:2001db}
  G.~Holder, Z.~Haiman, J.~Mohr,
  Astrophys.\ J.\  {\bf 560 } (2001)  L111-L114.
  [astro-ph/0105396].

\bibitem{Geisbuesch:2006fr}
  J.~Geisbuesch, M.~Hobson,
  Mon.\ Not.\ Roy.\ Astron.\ Soc.\  {\bf 382}, 158 (2007),
  [astro-ph/0611567].

\end{thebibliography}
\end{document}